\documentclass[preprint,column]{aastex}
\def\sss{\scriptscriptstyle}
\def\^#1{^{\sss #1}}
\def\_#1{_{\sss #1}}
\def\beq{\begin{equation}}
\def\eeqno#1{\label{#1}\end{equation}}

\def\rarrow{\rightarrow }

\def\dleft{\rlap{{\it D}}\raise 8pt\hbox{$\scriptscriptstyle\Leftarrow$}}
\def\dright{\rlap{{\it
D}}\raise 8pt\hbox{$\scriptscriptstyle\Rightarrow$}}

\def\au{{\rm au}}

\def\msun{M\_{\odot}}
\def\az{a\_{0}}

\def\l0{\ell\_{0}}

\def\rar{\rightarrow}
\def\s{\sigma}
\def\b{\beta}

\def\grcm{gr~cm^{-3}}

\def\l{\lambda}

\def\f{\phi}

\def\k{\kappa}

\def\r{\rho}
\def\rp{\rho_p}

\def\m{\mu}
\def\n{\nu}

\def\d{\delta}

\def\a{\alpha}

\def\xlimin{{x\rarrow\infty \atop{\raise 1pt\hbox to 30pt{\rightarrowfill}}}}
\def\limlim#1#2{{#1\rarrow #2 \atop{\raise 1pt\hbox to 30pt{\rightarrowfill}}}}

\def\vr{{\bf r}}

\def\vg{{\bf g}}

\def\vn{{\bf n}}

\def\S{\Sigma}

\def\grad{\vec\nabla}
\def\div{\vec \nabla\cdot}
\def\gf{\grad\phi}

\def\RM{r\_M}
\def\fN{\f\_N}
\def\gN{g\_N}
\def\gfN{\gf\_N}
\def\fpg{4\pi G}

\begin{document}

\title{A novel MOND effect in isolated high-acceleration systems}
\author{Mordehai Milgrom }
\affil{DPPA, Weizmann Institute}

\begin{abstract}
I discuss a novel MOND effect that entails a correction to the dynamics of isolated mass systems even when they are deep in the Newtonian regime: systems whose extent $R\ll \RM$, where $\RM\equiv (GM_t/\az)^{1/2}$ is the MOND radius, and $M_t$ the total mass. Interestingly, even if the MOND equations approach Newtonian dynamics arbitrarily fast at high accelerations, this correction decreases only as a power of $R/\RM$. The effect appears in formulations of MOND as modified gravity, governed by generalizations of the Poisson equation. The MOND correction to the potential is a quadrupole field $\f_a\approx G\hat Q_{ij}r^ir^j$, where $\vr$ is the radius from the center of mass. In QUMOND (quasilinear MOND), $\hat Q_{ij}=-\a Q_{ij}\RM^{-5}$, where $Q_{ij}$ is the quadrupole moment of the system, and $\a>0$ is a numerical factor that depends on the interpolating function. For example, the correction to the Newtonian force between two masses, $m$ and $M$, a distance $\ell$ apart ($\ell\ll \RM$) is $F_a=2\a(\ell/\RM)^3(mM)^2M_t^{-3}\az$ (attractive).
Its strength relative to the Newtonian force is $2\a(mM/M_t^2)(\az/\gN)^{5/2}$ ($\gN\equiv GM_t/\ell^2$).
For generic MOND theories, which approach Newtonian dynamics quickly for accelerations beyond $\az$, the predicted strength of the effect in the Solar system is rather much below present testing capabilities. In MOND theories that become Newtonian only beyond $\k\az$, the effect is enhanced by $\k^2$.

\end{abstract}

\keywords{}

\section{introduction}
One would suppose (as had I for many years) that MOND dynamics always approach Newtonian dynamics, in regions of high accelerations, as precipitously as its field equations approach those of Newtonian dynamics. (For a recent review of MOND see Famaey \& McGaugh 2012.) One would surmise that if the interpolating function that characterizes the theory, $\m(x)$, approaches unity, very quickly, at high $x$, then the field, the forces, the dynamics in general, approach their Newtonian values as quickly, when the system accelerations become high compared with the MOND constant $\az$.
This may be so in some formulations (e.g., in `modified-inertia' ones) and it is so for spherical systems, in general.
But, I show here that this is not the general rule in formulations of MOND as an extension of the Poisson equation: the nonlinear Poisson formulation of Bekenstein \& Milgrom (1984), and the Quasi-linear\footnote{Since the term `quasi-linear' partial differential equations is used by mathematicians in a different sense than here, I am now referring to such theories as `practically linear'} formulation (QUMOND) of Milgrom (2010a). No matter how vanishingly small $1-\m$ is for high $x$--namely how close the field equations themselves approach the Poisson equation at high accelerations--in aspherical systems, $g/\gN-1$ remains finite, and is of order $(\az/g)^{5/2}$; here $g$ and $\gN$ are the MOND and Newtonian accelerations, respectively.
\par
Previously noted MOND corrections in systems with $g\gg\az$, are of several distinct types:

(1) Corrections stemming from remaining departure of $\m(x)$ from unity even for high accelerations ($x\gg 1$). Such corrections have been discussed extensively since the early days of MOND (e.g., Milgrom 1983, Sereno \& Jetzer 2006, Milgrom 2009a). I do not discuss these here. To isolate away such effects, I shall assume here that $\m(x)-1\rar 0$ fast enough as $x\rar\infty$ (how fast will become clear below). We can then put $\m=1$ for $x\gg 1$, to the desired accuracy.

(2) Effects stemming from a MONDian background field in which the system is embedded, such as the galactic field in the context of a stellar system. This effect and its consequences for the Solar system have been considered in detail in Milgrom (2009a) and in Blanchet \& Novak (2011), and these too are not considered in this paper, where I assume that the system is isolated.

(3) Even in systems that are of high acceleration almost everywhere, there are, generically, small regions where the accelerations are smaller than $\az$. These include, e.g., the very centers of stars and the regions around the points of zero gravity in many-body systems. Such near-zero-gravity regions in the Solar system have been discussed as possible sites for testing MOND using test-particles probes (Bekenstein \& Magueijo 2006, Magueijo \& Mozaffari 2012, Galianni \& al. 2011).
Here I am not interested in the direct probing of such regions, but, rather, in the effect their presence has on the dynamics of the mass sources themselves. I shall show that the effects of these zero-gravity points are, generically, much smaller than the new effect I discuss here.
\par
In section \ref{qualitative}, I describe the new effect qualitatively. In section \ref{qumond}, I calculate it in the framework of QUMOND. In section \ref{bm}, I comment on the effect in the nonlinear-Poisson formulation. Section \ref{zero} treats the zero-gravity points in the system. In section \ref{discussion}, I mainly discuss the effect in theories that restore Newtonian dynamics only at accelerations much larger than $\az$.

\section{\label{qualitative}Qualitative explanation}
Consider an isolated distribution of nonrelativistic masses, $\r(\vr)$, of total mass $M_t$, contained within a region of radius $R$, much smaller than the MOND radius of the system, $\RM\equiv (M_t G/\az)^{1/2}$. Also assume that the system is quasi-static, in the sense that it varies only on time scales much longer than $\RM/c$.
Apart for small regions around zero-gravity point--which I consider separately in section \ref{zero}, and which have negligible effects--the accelerations in the system are everywhere much larger than $\az$, and hence we may put $\m=1$ everywhere within $R$. In such regions, the MOND equations coincide with the standard Poisson equation, and so the MOND potential satisfies the linear Poisson equation. However, it is not the standard, Newtonian solution, which assumes that the Poisson equation is valid everywhere to infinity. The fact that the field equation departs from Poisson beyond $\RM$, imposes a different solution even at much smaller radii.
\par
It is easy to understand the effect with the aid of the so called `phantom' mass (PM) distribution of the system, defined as (Milgrom 1986b)
 \beq \rp=\frac{1}{4\pi G}(\Delta\f-\Delta\fN),  \eeqno{i}
where $\f$ and $\fN$ are the MOND, and Newtonian, potentials of the system, respectively. In other words, the MOND correction to the Newtonian field is simply the Newtonian field of the PM.
This concept is very useful in MOND generally, where it helps us bring to bear our long experience with Newtonian gravity, especially in cases where we have some prior knowledge of the PM distribution. The concept of PM was used effectively, e.g., in Milgrom (1986b), Milgrom \& Sanders (2008), Wu \& al. (2008), and in Zhao \& Famaey (2010). It is  especially useful in the context of QUMOND, where $\rp$ can be easily calculated at the outset.
\par
Quasi-staticity, in the sense defined above, has to hold for the auxiliary of PM to be useful, since we assume that its distribution is uniquely determined by the instantaneous distribution of $\r$. Our nonrelativistic approximation breaks down if the system varies on time scales that are not much longer than $\RM/c$.

\par
Under the above conditions, the MOND dynamics of any high-acceleration, spherical system are exactly Newtonian. This follows by applying the Gauss theorem to spherical volumes concentric with the system. The fact that asymptotically the field crosses to the MOND regime is immaterial inside the system. Viewed differently, for such a spherical system with $R\ll\RM$, the phantom mass forms a hollow, spherical cavity roughly beyond $\RM$; so it does not affect the dynamics of the masses within the system, which remain strictly Newtonian.
\par
When, however, the system $\r$ is aspherical, the distribution of $\rp$ is also aspherical. So, even if it is hollow inside $\RM$, it can introduce finite MOND effects into the dynamics of our system. It is these effects that I want to calculate.
\par
Since $R\ll\RM$, the MOND field at, and beyond, $\RM$ is spherical to lowest order, with a correction of  quadrupolar angular distribution.\footnote{I assume that the quadrupole of the system does not vanish. Otherwise, the dominant correction is of a higher multipole, and of a higher order in $R/\RM$.} Then the PM is also spherical to lowest order, plus a small aspherical contribution. The spherical part has no effect on the dynamics of the system. Since the aspherical PM lies beyond $\RM$, its added potential much inside $\RM$ is of the quadrupole type (the dipole will be shown to vanish)
\beq \f_a\approx G\hat Q_{ij}r^ir^j.  \eeqno{mupa}
This is a vacuum solution of the Poisson equation since $\hat Q_{ij}$ is traceless.
It is $\hat Q_{ij}$ that I want to calculate.
\par
To recapitulate, much inside $\RM$ the MOND acceleration field is very high; so there $\m=1$ (apart from the above mentioned, inconsequential, MOND islands). Thus, $\f$ is, there, a solution of the Poisson equation. But it is not the standard solution, $\fN$, which is oblivious of the MOND limit of the theory beyond $\RM$. Instead, it is $\f\approx\fN+\f_a$, while at larger radii, $r\gtrsim\RM$, $\f$ becomes the full MOND solution of the problem.
Because of the assumption of quasi-staticity, we can assume that even if the system changes over time, the distribution
of PM `adjusts itself' continuously to the instantaneous configuration.
\subsection{Quasi-static approximation}
Even in deeply nonrelativistic systems, there are aspects that require relativistic treatment. Such is the case, e.g., when we consider influences over distances for which the light travel time is longer than time scales over which the influences vary. Such may be the case in the present context: The origin of the anomaly here is the behavior of the field at distances $\gtrsim \RM$, which, in turn, is determined by the matter distribution near the origin. If the latter changes on a time scale $\tau$ that is not much longer than $2\RM/c$, the field at $\RM$ cannot be assumed to adjust to the instantaneous configuration,
and instantaneously influence back the dynamics within $\r$. A relativistic treatment is then needed, and my treatment below is not valid.
\par
In a system, such as a binary, whose $\r$ varies on a dynamical time scale $\tau\sim R/v\sim (R^3/MG)^{1/2}$, the condition for quasi-staticity can be written as $4(\gN/\az)(v/c)^2\ll 1$.  This can also be written as
\beq R\gg R\_R\equiv (2R_s\RM^2)^{1/3}\sim (2\pi R_s^2\ell_H)^{1/3}\sim 1(M/\msun)^{2/3}\au \eeqno{musmus}
($R_s$ is the Schwarzschild radius of $M$, and $\ell_H$ the Hubble radius).
\par
When the opposite of inequality (\ref{musmus}) holds,
one might be tempted to simply time average the distribution of the PM; but this has to be justified via a relativistic treatment.

\section{\label{qumond}Calculation in QUMOND}
QUMOND (Milgrom 2010a) is a practically linear formulation of MOND derived from an action. It is the nonrelativistic limit of a certain formulation of bimetric MOND (BIMOND) (Milgrom 2009b).
The field equation for the gravitational potential is
 \beq \Delta\f=\div[\nu\left(\frac{|\gfN|}{
\az}\right)\gfN],~~~~ {\rm where}~~~~~~~~\Delta\fN=\fpg\r, \eeqno{ii}
with $\gf\rar 0$ at infinity.
Here, $\n(y\rar\infty)\rar 1$, and $\nu(y\rar 0)\rar y^{-1/2}$. It is related to the usual MOND interpolating function, $\m(x)$, by $\n[x\m(x)]=1/\m(x)$.
QUMOND thus requires solving (twice) the linear Poisson equation (and not a nonlinear differential equation). It has already been put to good use for predicting and calculating
MOND effects in the Solar system (Milgrom 2009a, Galianni \& al. 2011), for calculating MOND fields of galaxies (e.g. Angus et al. 2012), and for structure formation in MOND (Llinares 2011); see also Zhao \& Famaey (2010). Its relation to the nonlinear-Poisson formulation is elaborated on in Milgrom (2012). For example, it is shown there that the two theories are equivalent  for one-dimensional systems when the nonlinear Lagrangians of the two theories are the Legendre transforms of each other, leading to the above relation between $\m(x)$ and $\n(y)$.
\par
In regions where $\r=0$, the phantom density is given by
 \beq \rp=\frac{1}{\fpg}\div[\nu\left(\frac{|\gfN|}{
\az}\right)\gfN], \eeqno{nutreq}
and is easily calculated from the Newtonian potential.
Since $\rp$ is nonzero only at distances much larger then $R$, we can take only the lowest contributing harmonic to the Newtonian potential. Taking the origin at the center of mass of the system, the dipole moment vanishes, and we keep only the monopole and quadrupole contributions, writing
 \beq \fN\approx -\frac{GM_t}{r}+\eta, \eeqno{milas}
where,
 \beq  \eta=Gr^{-5}r^ir^j Q_{ij},~~~~~~~~~~~Q_{ij}=\frac{1}{2}\int \r(\vr')(r'^2\d_{ij}-3r'_i r'_j)d^3r'. \eeqno{solga}
\par
Using eqs.(\ref{milas}-\ref{solga}) in expression (\ref{nutreq}), keeping only terms up to first order in $\eta$, we get:\footnote{All the first-order terms are scalars that are linear in $Q_{ij}$, so they must be of the form $Gf(r)r^ir^j Q_{ij}=f(r)r^5\eta$ ($Q_{ii}=0$). For example, use is made of $\vr\cdot\grad\eta=-3\eta$, and $r^ir^j\eta_{,i,j}=12\eta$.}
 \beq\fpg\rp\approx  -2y\n'(y)\frac{GM_t}{r^3}
 +[18y\n'(y)+6y^2\n''(y)]\frac{\eta}{r^2},
     \eeqno{ropa}
where, $y=(\RM/r)^2$.
The first term is spherical. At large distances, where $y\ll 1$, so $\n'\rar -y^{-3/2}/2$, it gives the asymptotic phantom matter `isothermal halo': $\rp \propto M_t/\RM r^2$. At short distances, $\n'\rar 0$, and, as stated above, I assume that it does so fast enough that we can take it as zero within $R$. The first term thus describes a spherical distribution with a spherical cavity; so it has no effect on the dynamics of $\r$. In the second term, call it $\fpg\hat\rp$, all the factors are spherical, except $\eta$, which is a quadrupole. This part of $\rp$ also has an empty cavity for $r\ll\RM$, but it does produce a field inside it. Because $\hat \rp$ is nonvanishing only at distances $r\gg  R$, we can keep only its lowest multipole contribution, which is the quadrupole, since $\eta$ is reflection symmetric. Its field within $R$ can be written to the dominant power in $r/\RM$ as
\beq \f_a\approx Gr^ir^j\hat Q_{ij}, ~~~~{\rm where}~~~~\hat Q_{ij}=\frac{1}{2}
\int \hat\rp(\vr')(r'^2\d_{ij}-3r'_ir'_j)\frac{d^3r'}{r'^5}. \eeqno{sholas}
Inserting expression (\ref{ropa}) for $\hat\rp$, with expression (\ref{solga}) for $\eta$, we get
 \beq \hat  Q_{ij}=\frac{9}{8\pi \RM^5}Q_{kl}\int_0^{\infty}dyy^{5/2}[\n'(y)
+\frac{y}{3}\n''(y)]
 \int d\Omega(\d_{ij}-3n_in_j)n_kn_l,  \eeqno{miobes}
where $\vn=\vr'/r'$. For the radial integral to converge at $r\rar 0$, it is necessary that $\n'(y)$ vanishes at large $y$ faster than $y^{-7/2}$. This is, indeed, what is quantitatively meant when I say repeatedly that $\n$ is assumed to approach 1 fast enough. For slower vanishing, the MOND correction of type 1 discussed in the introduction can be shown to dominate the present effect, and our approximation here is not valid.
\par
After performing the angular integral, and integrating the $y$ integrals by parts, we get
\beq \hat  Q_{ij}=-\frac{\a}{\RM^5}Q_{ij},~~~{\rm where}~~~\a=\frac{3}{4}
\int_0^{\infty}dyy^{3/2}[\n(y)-1].  \eeqno{mioop}
Thus, $\a$ is a numerical factor that depends on the interpolating function.\footnote{Note that $\hat Q_{ij}$ is the `internal' quadrupole of the PM. It is finite, and picks up its main contribution from radii around $\RM$. The expression for the `external' quadrupole moment of the PM, defined as in eq.(\ref{solga}), diverges linearly at large radii (small $y$) for isolated systems, just as the total PM mass does. This does not lead to divergences in the fields, and is, anyhow, cut off by external fields. In some systems there are also departures from quasi-staticity to be reckoned with at large radii.}
\par
A generic choice of $\n(y)$ in MOND would be one for which $\n(y)-1$ is $\approx y^{-1/2}$ for $y\ll 1$, and can be made to vanish quickly for $y\gg 1$. Such a choice would give $\a\sim 1$, to within an order of magnitude roughly.\footnote{Of course, $\a$ can be made arbitrarily large if $\n-1$ vanishes nearly as $y^{-5/2}$.}
For example, for the very sharply transiting, limiting form of $\m$: $\m(x)=x$ for $x\le 1$, and $\m(x)=1$ for $x\ge 1$ [for which $\n(y)=y^{-1/2}$ for $y\le 1$, and $\n(y)=1$ for $y\ge 1$], we get $\a=3/40$. For $\n(y)=(1-{\rm e}^{-y})^{-1/2}$ one gets $\a\approx 0.6$. For the slowly transiting $\n(y)=(1-{\rm e}^{-y^{1/2}})^{-1}$, which was used successfully in Famaey \& McGaugh (2012) for rotation-curve predictions, we have $\a\approx 37$.
\par
There may also be interest in interpolating functions for which very-near-Newtonian behavior is reached, not beyond $\gN\sim\az$, but beyond $\gN\sim\k\az$, with, possibly, $\k\gg 1$ (see section \ref{discussion}). For these,
$\n-1$ is made to vanish quickly only for $y\gg \k$, while for $y\ll\k$, $\n-1\approx y^{-1/2}$.
We can write some such functions in the form
\beq \n=\n_{\k}(y)=1+\k^{-1/2}[\n(y/\k)-1], \eeqno{mavy}
where $\n$ is of the generic type defined above. Substituting expression (\ref{mavy}) in eq.(\ref{mioop}), we see that for such a function $\a_{\k}=\k^2\a_{\k=1}$, where $\a_{\k=1}$ is roughly of order 1.

\par
For a highly aspherical system, for which $Q\sim M_t R^2$ (such as a system of two comparable masses), the anomalous acceleration is $g_a\sim GM_tR^3/\RM^5$, and its strength relative to the Newtonian acceleration, $\gN\sim GM_t/R^2$, is $g_a/\gN\sim (R/\RM)^5\sim (\az/\gN)^{5/2}$. This is so, I reemphasize, even if $\n(\gN/\az)-1$ approaches zero arbitrarily fast with increasing $\gN/\az$.
In almost spherical systems, the anomaly is further reduced. If  $m\ll M_t$ is the `aspherical' mass (e.g., the mass of planets in a planetary system) then the anomalous MOND acceleration in the system is $g\_a\sim (m/M_t)(R/\RM)^3\az$.
\par
As an example, consider the MOND correction to the Newtonian, two-body force between masses $m$ and $M$, a distance $\ell$ apart, for which condition (\ref{musmus}) for quasi-staticity holds. Here,
$Q_{zz}=-\ell^2mM/M_t$, and $Q_{xx}=Q_{yy}=-Q_{zz}/2$, where the masses lie on the $z$ axis. The MOND correction to the Newtonian force is then attractive, and equals
\beq F_a=2\a(\ell/\RM)^3(mM)^2M_t^{-3}\az.  \eeqno{kinla}
\section{\label{bm}The nonlinear-Poisson theory}
In this theory, the MOND potential for an isolated system is the (unique) solution of the equation (Bekenstein \& Milgrom 1984)
\beq
\div[\m\left(\frac{|\gf|}{\az}\right)\gf]=\fpg\r,\eeqno{xii}
with $\gf\rar 0$ at infinity. This is the nonrelativistic limit of
Einstein-Aether theories (Zlosnik \& al. 2007),  and it is also part of the nonrelativistic limit of TeVeS (Bekenstein 2004), and of generic formulations of BIMOND (Milgrom 2009b, 2010b).

If $\m$ is related to $\n(y)$ of QUMOND as described above, the two theories coincide for spherical systems, and the (radial) dominant asymptotic behavior is the same in the two theories.
Again, in the phantom-density approach, we need the dominant, asymptotic, aspherical MOND potential. However, in this theory we cannot calculate $\rp$ before a full solution of the problem is known.
\par
In Milgrom (1986a), I showed that the aspherical, far field still has, generically, a quadrupolar angular dependence, but its radial dependence is somewhat different from that in QUMOND.
The asymptotic potential, beyond $\RM$, is of the form
\beq \f\approx (G\az M_t)^{1/2}ln(r) +G\frac{S_{ij}r^ir^j}{r^{(2+\sqrt{3})}},\eeqno{loca}
where $S_{ij}$ is a symmetric, traceless (constant) matrix; it depends on the mass distribution $\r$, but I do not know how.
The phantom density outside $\r$ is $\fpg\rp=\Delta \f$, so its asymptotic form is
\beq \fpg\rp\rar \frac{GM_t}{\RM r^2}+  \b G\frac{S_{ij}r^ir^j}{r^{(4+\sqrt{3})}}, \eeqno{sasat}
where $\b=(\sqrt{3}+2)(\sqrt{3}-3)$.
Here too, in high-acceleration regions, roughly within $\RM$, $\rp$ vanishes. This is because there $\m=1$, so $\Delta\f=\Delta\fN$.
The first term in the expression for the asymptotic $\rp$ is spherical, and does not affect the dynamics in the system. The second term gives rise again to a quadrupole field
\beq \f_a\propto G\frac{S_{ij}r^ir^j}{\RM^{(3+\sqrt{3})}}. \eeqno{milta}
$S$ has the dimensions of $mass\times length^{\sqrt{3}}$. It vanishes for a spherical system, and also in the limit $\az\rar 0$. So, we may write it as $S\sim M_t R^{\sqrt{3}}\xi(R/\RM)$, where $R$ is the characteristic extent of the system. The dimensionless function $\xi(u)$ is unknown, however, and depends on the various dimensionless parameters of the system (mass ratios, ratios of distances, etc.), and has to vanish for $u=0$, which corresponds to $\az=0$.
Numerical calculations are needed to say more on this.

\section{\label{zero}Contribution of the small MOND domains near zero-gravity points}
Clearly, in every mass system there are critical points of the potential, where its gradient vanishes.\footnote{I speak here of the nonrelativistic potential, therefore barring singularities, such as black holes.} This follows from topological considerations, but is otherwise obvious. There are, for example, the points at (or very near) the centers of spherical stars or planets. Also, in a system of many compact objects, there are zero-gravity points, around which there is a MONDian region where $|\gf|<\az$.
These regions are surrounded by Newtonian regions, where we assume that $\m=\n=1$ to the desired accuracy. This implies that the total phantom mass within each such MONDian region vanishes: Applying the Gauss theorem to the volume $V$, within a surface $\Sigma$ that is wholly in the Newtonian region, we have $\int\_V\rp dV\propto \int\_\S(\gf-\grad\fN)\cdot d\vec\s=\int\_\S[\m(|\gf|/\az)\gf-\grad\fN]\cdot d\vec\s=\int_V\div[\m(|\gf|/\az)\gf-\grad\fN]dV=0$, where I used the fact that on $\Sigma$, $\m=1$.
In a similar way it is seen that this is also true in QUMOND.
\par
The effects of such regions on the dynamics of $\r$
can be calculated if we know the PM distribution in them.
Since the typical acceleration within such a region is, by definition, $|\gf|\sim\az$ (I restrict myself to MOND theories that approach Newtonian dynamics quickly beyond $\az$),
the characteristic size, $L$, and phantom density are related by $\fpg \rp L\sim \az$.
\par
The MONDian regions near the centers of spherical objects are, themselves, spherical (they may be slightly aspherical if the general geometry is strongly aspherical, but I neglect this); so, since they have vanishing total PM, they have no outside effect.\footnote{For a star of central density $\r$, the characteristic phantom density in the region is also $\sim\r$; therefore, the size of this region is $L\sim \az/G\r$, which for $\r\sim 1\grcm$, is a fraction of a centimetre.}

\par
Other zero-gravity points typically occur in vacuum, so they cannot be extrema of the potential; they are thus saddle points.
Take for example a system of a few comparable masses with inter-mass distances $\sim R$. If $\vg(\vr)$ is the gravitational acceleration field in the system, $g\sim GM_t/R^2$, then at a zero gravity point, $\vr_0$, we have $\vg(\vr_0)=0$.
The size of the region around this point, within which $g\lesssim\az$ is
$L\sim \az(dg/dr)^{-1}\sim (\az/g)R\sim R^3/\RM^2$.\footnote{If some masses are much smaller than others, then the zero-gravity points are much nearer such masses than $R$; say a distance $R_m$. Then $L\sim (\az/g)R_m$; but our result below remains valid.}
The total PM in this region vanishes. To lowest order in $R/\RM$ the region is reflection symmetric; so its dipole moment vanishes to this order (it would, otherwise scale as $\rp L^4$). The dipole moment is thus of the higher order: $d\sim \rp L^5/R$. The phantom quadrupole moment is $Q\sim \r L^5$. The acceleration field of the dipole at a characteristic distance $R$,
is $g_d\sim Gd/R^3\sim G\rp L^5/R^4$, and that of the quadrupole is $g\_Q\sim GQ/R^4\sim G\rp L^5/R^4$. They thus scale in the same way.
Since $GL\rp\sim \az$, we find that the anomalous MOND correction to the acceleration in the system is $g_a\sim \az (L/R)^4\sim \az(R/\RM)^8$.
This estimate can also be made for the case of a small mass $m$ in a larger, Newtonian system whose characteristic size is $R$, and whose total mass is $M_t\gg m$. Again, one estimates that the MOND correction to the acceleration on $m$ is $g_a\sim\az(R/\RM)^8$.
All this is confirmed by explicit calculations in QUMOND, which I omit here.
\par
The accelerations due to this effect thus scale as $(R/\RM)^8$, compared with $(R/\RM)^3$
for the correction discussed above; but for this effect the anomalous acceleration on a small mass does not vanish in the limit $m\rar 0$.
\section{\label{discussion}Discussion}
Intuitively, one would expect that the MONDian, two-body force  becomes exactly Newtonian at high accelerations, if the interpolating function does so. I have shown that this is not the case, and that, more generally, there are MOND effects that linger even in high-acceleration systems.
\par
For theories with $\a\sim 1$, I cannot think, at present, of a precision experiment to test this effect: in the Solar system, the effect is too small to be tested with present capabilities.
For example, for the dynamics of the massive planets, $m/M_t\sim 10^{-3}$, and $r\sim R\sim 10^{-3}\RM$; so the added MOND accelerations are of order $10^{-12}\a \az$, where present accuracy hardly reaches $\sim 10^{-4}\az$.
\par
Note that my treatment here is not applicable, as is, to short-orbital-period pulsar binaries, as they are anything but quasi-static: For binaries that serve as laboratories to test relativistic theories, having orbital periods of a fraction of a day, the orbital light cylinder radius is of the order of a light hour, much smaller than the MOND radius of these systems, which is larger than $10^3$ light hours. For the Solar system, we see that condition (\ref{musmus}) is satisfied by the large planets, which determine the quadrupole of the PM.
\par
But do we necessarily have $\a\sim 1$? In considering MOND, I have always labored under the supposition that it involves only one acceleration constant that plays all possible roles of such a constant. This, to me, is a basic tenet of MOND. This supposition is amply supported by rotation-curve and other dynamical studies of galaxies, where indeed the same acceleration constant appears with different roles (see, e.g., Famaey \& McGaugh 2012).\footnote{This is similar to $\hbar$ playing all the roles of an action constant in quantum mechanics, or to $c$ playing all the roles of a critical speed in relativity.}
In particular, this means that $\az$ that appears in the mass-asymptotic-speed relation, $MG\az=V^4\_{\infty}$, also marks the boundary of the Newtonian regime; namely, that beyond $\sim \az$, Newtonian behavior is reached quickly. This underlies my conclusion above regarding the Solar system, since it implies that $\a\sim 1$ within roughly an order of magnitude.
\par
This supposition holds (or can easily be made to hold) in most relativistic formulations of MOND:  MOND adaptations of Einstein-Aether theories (Zlosnik \& al. 2007),  BIMOND (Milgrom 2009b, 2010b), theories based on a polarizable medium (Blanchet \& Le Tiec 2009), and nonlocal, metric theories (Deffayet \& al. 2011).
However, it does not hold in TeVeS, in its original form. The basic reason behind this is that this form of TeVeS does not tend to general relativity (GR) in the formal limit $\az\rar 0$: it is not quite GR compatible. To get GR in such a TeVeS version, one also has to take a limit $k\rar 0$, where $k$ is a dimensionless parameter of TeVeS (possibly representing several such parameters in different TeVeS formulations). Since various tests, in the Solar syste and pulsar binaries, strongly constrain high-acceleration departures from GR, they force one to take $k\ll 1$. Now, in the nonrelativistic limit of such theories,
the fully Newtonian regime can occur only beyond $\gN\sim\az^*\equiv \k\az$, where $\k\gg 1$. Effectively, we then have two very different acceleration constants, $\az$ and $\az^*$, playing important roles in the theory. For example, Bekenstein \& Magueijo (2006) found that for the TeVeS formulation they used, the boundary of MOND regions around zero-gravity points are defined using $\k=(4\pi/k)^2\gg 1$ ($\k\approx 1.75\cdot 10^5$ for their choice of $k=0.03$). Starting from the same assumptions, Magueijo \& Mozaffari 2012 used even higher values of $\k\gtrsim 1.6\cdot 10^6$ (based on $k\lesssim 0.01$). Galianni \& al. 2011 follow a similar line in most of their paper, but do make the point that other, no less sensible, choices of a MOND theory, having $\k\sim 1$, give effects that are many orders smaller.
\par
In my opinion, a strong desideratum of a relativistic MOND theory that involves only one dimensioned parameter, $\az$, is GR compatibility: the theory should tend to GR in the formal limit $\az\rar 0$. This is analogous to the requirement that GR go to Newtonian dynamics for $c\rar\infty$; it is also analogous to Bohr's correspondence principle in quantum mechanics. All relativistic MOND theories mentioned above, except TeVeS, satisfy this requirement. Thus, in all these theories $\k\sim 1$--which is preferred also by galaxy dynamics--is admissible. As already mentioned, the original versions of TeVeS are forced to adopt $\k\gg 1$ because they are not GR compatible, a fact that puts them in danger of conflicts with tests in the high acceleration regime, unless $k\ll 1$. Babichev \& al. (2011) have recently devised a theory that might be viewed as a short-distance modification of TeVeS (introducing an additional scale length besides $\az$), and that avoids conflicts with observations, while admitting $\k\sim 1$.
I feel that {\it a theory with $\k\gg 1$, and all the consequences that follow from it, are not generic MOND results}.
\par
The QUMOND versions of theories with $\k\gg 1$ would involve an interpolating function as given in eq.(\ref{mavy}). Then, as we saw in section \ref{qumond}, the coefficient $\a$ defined in eq.(\ref{mioop}) is $\a=\a_{\k}= \k^2\a_{\k=1}$, where $\a_{\k=1}$ is of order 1, to within an order of magnitude, roughly. The high $\k$ values used in the above-mentioned analyses, could already be in conflict with existing Solar-system limits on the effect I discuss here. Checking this requires a detailed analysis. Here I give only a rough estimate: Assume that the quadrupole moment of the Solar system comes from
a single, most dominant planet, of mass $m$ and semi-major axis $\ell$. For convenience of comparison, write the anomalous potential as
 \beq \f_a=-\frac{Q_2}{2}r^ir^j(e_ie_j-\frac{1}{3}\d_{ij}). \eeqno{namana}
Then, from all the above, we have
 \beq Q_2\approx -3\k^2\a_{\k=1}G\ell^2m\RM^{-5}.  \eeqno{umalu}
Taking the dominant contribution from Neptune (Jupiter, Saturn, and Uranus give only somewhat smaller contributions), we get
 \beq Q_2\approx -3\cdot 10^{-34}\k^2\a_{\k=1}s^{-2}. \eeqno{lalala}
In comparison, the external-field effect studied by Milgrom (2009a), and by Blanchet \& Novak (2011), produces $Q_2\sim (0.2-4)\cdot 10^{-26}s^{-2}$. So, if $\a_{\k=1}\sim 1$, the effect discussed here is larger for $\k\gtrsim 10^4$. Values of $\k>10^5$ might already be excluded.
A more detailed analysis should consider that several planets contribute to the quadrupole of the Solar system, that these
moments are rotating with the planets, that our results were derived in QUMOND, while the nonrelativistic limit of a theory like TeVeS is not QUMOND, that the Solar system is not isolated, but embedded in the Galactic field, as well as other factors.

\clearpage

\end{document}